\documentclass[twocolumn,showpacs,preprintnumbers,amsmath,amssymb]{revtex4}
\usepackage{graphicx}
\usepackage{dcolumn}
\usepackage{bm}
\def\bchi{\bm{\chi}}
\def\bLambda{\bm{\Lambda}}

\begin{document}
\title{Modified Thouless-Anderson-Palmer equations for the
 Sherrington-Kirkpatrick spin glass:
Numerical solutions}
\author{T. Plefka}
\email{timm@arnold.fkp.physik.tu-darmstadt.de}
\affiliation{Theoretische Festk\"orperphysik, TU Darmstadt, D
64289 Darmstadt, Germany}
\date{\today}
\begin{abstract}
For large but finite systems  the static properties  of the
infinite ranged Sherrington-Kirkpatrick model are numerically
investigated in the entire  the  glass regime. The  approach is
based on the modified Thouless-Anderson-Palmer equations in
combination with a phenomenological relaxational dynamics used as
a numerical tool. For all temperatures and all bond configurations
stable and meta stable states are found. Following  a discussion
of the finite size effects, the static properties of the state of
lowest free energy are presented in the presence of a homogeneous
magnetic field for all temperatures below the spin glass
temperature.  Moreover some characteristic features of the meta
stable states are presented. These states exist in finite
temperature intervals and disappear via local saddle node
bifurcations. Numerical evidence is found that the excess free
energy of the meta stable states remains finite in the
thermodynamic limit. This implies  a the `multi-valley' structure
of the free energy  on a sub-extensive scale.
\end{abstract}
\pacs{75.10.Nr, 05.50.+q, 87.10.+e}
\maketitle
\section{Introduction}
The Sherrington-Kirkpatrick (SK) model \cite{sk} for Ising spins
with  quenched random bonds is the simplest representative  of a
class of long-ranged models all describing successfully the
interesting phenomena of spin glasses \cite{mpv,fh}. In addition
to this success  in physical questions, the research on these
models has been fruitful and stimulating in optimization problems,
in the understanding of neural networks and in the field of
learning machines \cite{mpv,her}.

Two complimentary but conceptually different approaches exist for
the analysis of these models. The first approach is the replica
method  applied already by Sherrington and Kirkpatrick to their
model in the original work. After breaking the replica symmetry it
was Parisi \cite{par}  who first found a satisfactory  solution of
the SK model in the spin glass regime.

Within the replica theory and the related dynamical approaches
\cite{sz},  an averaging over the quenched variables is performed
and therefore only the calculation of bond-averaged quantities is
possible. All  problems  that depend on a special, but arbitrary
configuration of the random bonds cannot be solved within the
framework of the replica theory. Such problems are the generic
case in the nonphysical applications named above. In optimization
problems, for instance, the main question is to find the optimal
solution for a fixed configuration.

The complimentary approach of Thouless-Anderson-Palmer (TAP)
\cite{tap} to investigate spin glass models does not perform the
bond average  and  permits the treatment of problems depending on
specific configurations. For other questions which are  expected
to be independent of the special configuration, such as all
macroscopic physical quantities, self averaging happens. This is
due to the fact that the random interaction matrices have well
known asymptotic properties in the thermodynamic limit \cite{ma}.
The situation is in principle similar to the central limit theorem
in probability theory, where large numbers of random variables
also permit the calculation of macroscopic quantities which hold
for nearly every realization of the random variables. Thus the
investigation of one or some representative systems is sufficient
and the bond average is not needed.

 The TAP equations have been well
established for more than two decades and several alternative
derivations are known \cite{mpv,fh,I}. Nevertheless the TAP
approach is still a field of current interest. This is due to the
importance of the approach to numerous interesting problems.
Moreover it is suspected that not all aspects of this approach
have yet been worked out. Very recently the author \cite{II} has
reanalyzed the stability of the TAP equations. He concluded that
the reason for the breakdown at the spin glass instability is an
inconsistency in the value of the local susceptibility and showed
that unstable states cannot be described by the original TAP
equations. Removing these inconsistencies  by a self-consistent
treatment, modified TAP equations below the instability have been
presented \cite{II}.

One of the  pressing questions  is the characteristic features of
the pure states of the SK model, which are the stable  solutions
of the TAP equations.  It is obvious that these states are
essential for a direct understanding of the SK model. Moreover the
pure states are of importance for the interpretation of the more
formal results of the replica theory \cite{mpv}.

Apart from the research \cite{num}, showing that the number of the
pure states is very large  at low temperatures, two approaches
exist to investigate the characteristic features of the pure
states. Both the work of Bray and Moore \cite{bm} and the work of
Nemoto and Takayama \cite{nt} try to find explicitly  numerical
solutions of the TAP equations for large but finite systems. Only
in rare cases (about 15 \%  of the systems with different bond
configurations) do these investigations yield solutions of the TAP
equations which satisfy the validity condition. To treat the other
cases Nemoto and Takayama have developed an interesting but
indirect method to obtain at least approximative solutions which
are expected to converge to true solutions in the thermodynamic
limit. The applied numerical method of \cite{nt} did not work well
for a cooling of the sample and therefore  results for this case
have not been published.

In this work we will demonstrate that all the known difficulties
to find numerically the pure states disappear if the modified TAP
equations \cite{II} are applied instead of the original ones. For
all configurations of the bonds, stable solutions of these TAP
equations will be found for finite systems employing a simple
relaxational dynamics. There is no further need to apply the
indirect method of \cite{nt} and our approach permits the sample
to be cooled down, as is shown to be essential to find the low
lying states. In addition to these general goals some
characteristic features of pure  states will be worked out leading
to a better understanding of the SK spin glass.

In Section II the basics of the modified TAP equations, some
essentials of the applied dynamics and the numerical procedure are
presented. A main subject of section III is the investigation of
finite size effects which are found to be of some importance.  The
numerical results of the state of lowest free energy  for  the
relevant static quantities will be given for all temperatures
below the critical spin glass temperature. Next we focus on the
meta stable states and discuss some characteristic features.
Finally, the concluding remarks can be found in section IV.
\section{ Basic equations and method}
\subsection{The modified TAP equations}
The Hamiltonian of the SK model of a system of N Ising spins
($S_i=\pm1)$ in the presence of local external fields $h_i$ is
described by
\begin{equation}\label{1}
  H= -\frac{1}{2} \sum_{i\neq j} J_{i j} S_i S_j -\sum_i h_i
  S_i\quad .
\end{equation}
The bonds $ J_{ij}$ are independent random variables with zero
means and standard deviations $ N^{-1/2}$  (where the latter
scaling fixes the spin glass temperature to $T=1$ ).

The modified TAP mean field equations \cite{II} describe the
statics of the system (\ref{1}) in the thermodynamic limit. They
are given  by the set of equations for the local magnetizations ,
with $m_i=\langle S_i\rangle_\beta\, $
\begin{equation}\label{4}
 m_i=\tanh \beta\big\{h_i +\sum_j J_{ij}m_j-m_i \chi_l\big\}
\end{equation}
together with the local susceptibility
\begin{equation}\label{7}
\chi_l =\frac{1}{N}\sum_i
\frac{\beta(1-m_i^2)}{1+\Gamma^2\,\beta^{2}(1-m_i^2)^2}
\end{equation}
and with
\begin{eqnarray}\label{8q}
\Gamma &=& 0\quad \hspace{4.2cm}\mathrm{for}\,x\geq 0\\
1&=&\frac{1}{N}\sum_i
\frac{\beta^2(1-m_i^2)^2}{1+\Gamma^2\,\beta^{2}(1-m_i^2)^2}\qquad
 \mathrm{for}\,x\leq 0\label{8neu}
\end{eqnarray}
where
\begin{equation}\label{5}
x= 1-\beta^2( 1-2q_2+q_4)
\end{equation}
and where
\begin{equation}\label{3}
q_\nu=N^{-1}\sum_i m_i^\nu \qquad \nu=2,4
\end{equation}
was introduced.

The condition $ x=0$ represents the central spin glass instability
condition \cite{at,mpv,fh}. Above the instability the local
susceptibility $ \chi_l$  reduces to the isothermal value
$\chi_l^{(\beta)}$
\begin{equation}\label{6}
\chi_l^+=\chi_l^{(\beta)}\equiv \beta( 1-q_2)\qquad \textrm{for
}x\geq0
\end{equation}
which is in complete agreement with the original TAP approach
\cite{tap,I,mpv,fh}. Essential differences, however, result below
the instability as
\begin{equation}\label{xxx}
\Gamma^->0\;,\qquad\chi_l^-\neq\chi_l^{(\beta)}\qquad \textrm{for
}x<0
\end{equation}
holds for the modified TAP equations. According to \cite{II} the
quantity $\Gamma$ is proportional to the density of zero
eigenvalues of the inverse  susceptibility matrix. Thus all the
states with $ x<0$ are unstable. As thermodynamics does not apply
to unstable states, a conflict with the exact thermodynamic
relation $ \chi_l=\chi_l^{(\beta)}$  does not exist.

As worked out in \cite{II} the modified Tap equations result from
the simple fact that the eqs.(\ref{4}) together with the
definition $  \chi_l=N^{-1} \sum_i \partial m_i/\partial h_i$ are
complete and sufficient to determine the $m_i$. From the strict
thermodynamic point of view the modified equations are equivalent
to the original TAP equations as thermodynamics is a priori
limited to stable states. Note, however, that unstable states are
essential for discussions based on free energy landscapes or for
the extension to dynamics of the present approach. For a further
discussion of this point we refer to \cite{II} and to the end of
the next subsection.

For later use we give the well known expressions
\cite{tap,I,mpv,fh} for the energy
\begin{equation}\label{81}
U=-\frac{1}{2}\sum_{i\neq j}J_{ij}m_i
m_j-\frac{\beta}{2} N(1-q_2)^2-\sum_i h_im_i
\end{equation}
and the entropy
\begin{eqnarray}\label{82}
 S=&- &\sum_i \Big\{\frac{1+m_i}{2}\ln
\frac{1+m_i}{2}+\frac{1-m_i}{2} \ln \frac{1-m_i}{2}\Big\}\nonumber\\
&-&N\frac{\beta^2}{4} N(1-q_2)^2
\end{eqnarray}
from which the free energy
\begin{equation}\label{83}
F=U-T S
\end{equation}
can be calculated. Note that  these expressions  were obtained by
thermodynamic approaches. Hence they are restricted to stable
states.

\subsection{Glauber dynamics}
As in \cite{dyn,fh}, Glauber dynamics in mean field approximation
will phenomenologically   be added  to the model. Measuring the
time $t$ in units of the relaxation time, the purely relaxational
equations of motion are given by
\begin{equation}\label{9}
 \dot{m}_i(t) = - m_i+ \tanh \beta \big \{h_i +\sum_j
J_{ij}m_j-m_i \chi_l (t)\big \}.
\end{equation}
The local susceptibility $ \chi_l(t)$ is related to the $ m_i(t)$
via  eq.(\ref{7}) and via eq.(\ref{8q})  or eq.(\ref{8neu})
depending on the instantaneous value of $ x(t)$. The fixpoints of
these equations of motion obviously coincide with the solutions of
the modified TAP equations.

To analyze  the dynamical stability of these fixpoints  the
equations of motion are linearized near a fixpoint ${m_i}$.
Setting
\begin{equation}\label{10}
\delta m_i(t)= m_i(t)-m_i
\end{equation}
and neglecting as usual the $ N^{-1}$ order terms $
\propto\partial\chi_l/\partial m_i$ , the linearized equations
take in matrix notation the form
\begin{equation}\label{11}
\delta\dot{m}= -\bchi_0 \bchi ^{-1}\delta m
\end{equation}
where $\bchi_0 $ represents the static susceptibility matrix for
noninteracting Ising spins
\begin{equation}\label{12}
(\chi_0)_{ij}=\beta( 1-m_i^2)\;\delta_{ij}
\end{equation}
and where
\begin{equation}\label{13}
\bchi^{-1}=\bchi^{-1}_0 +\chi_l \mathbf{1}-\mathbf{J}
\end{equation}
is the inverse of the  susceptibility matrix of the modified TAP
approach \cite{II}. For thermodynamic stable or marginal stable
states $\bchi^{-1}$ is positive semi-definite.

According to eq.(\ref{11}) the dynamical stability is governed the
matrix $\bLambda=\bchi_0 \bchi ^{-1} $. Let $ l $ be an
eigenvector of $\bLambda $ satisfying  the eigenvalue equation
$\bLambda \,l= \lambda \,l$. Multiplication with $\bchi^{-1}_0$
leads to $\bchi^{-1} \,l= \lambda\; \bchi^{-1}_0\,l $ and to
\[\lambda=\frac{l\cdot\bchi^{-1}\,l}{l\cdot\bchi^{-1}_0\,l}\:.
\]
Immediately  $ \lambda\geq0$ results  from the definiteness of
$\bchi_0$ and of $\bchi^{-1}$ for stable (or meta stable) static
states. Static stability  therefore implies dynamical stability.
By a similar argument, the statement can be proven in the opposite
direction, leading to a one-to-one correspondence between the two
types of stability.

The latter result implies that the stable solutions of the TAP
equations can be found by integration of the equations of motions
(\ref{9}). The system relaxes to a stable fixpoint which is
certainly  located outside the regime $x<0$. For transient times,
however, the system can enter  the regime $x(t)< 0$ and thus the
flow of eqs.(\ref{9}) is needed in both regions $x(t)<0$ and
$x(t)>0$. These arguments demonstrate the relevance of the
unstable states and consequently the need for the modified TAP
equations for the present approach. It should be added that
equations of motion of the above type but based on the original
TAP equations usually lead to incorrect results in the spin glass
region. For this case the system generally relaxes to paramagnetic
solutions.

\subsection{Systems of finite size}
The modified TAP equations are exact in the thermodynamic limit
but are approximative for systems of finite size. As numerical
studies can only be performed for finite systems  the two
approximations yielding  to eqs.(\ref{4}-\ref{8neu}) for
\textit{finite} $N$ are recalled.

First of all eqs.(\ref{4}) holds only to leading order in $ N^{-1}
$ as sub-extensive contributions have been neglected (compare e.g.
appendix of \cite{II}). Further corrections  to eq.(\ref{8neu})
result from the application \cite{II} of the Pastur theorem
\cite{pastur} which is again  only exact in the thermodynamic
limit. The explicit form of these two types of corrections are not
known and seem to be hard to determine analytically. Hence the
finite size effects resulting from the two approximations will be
investigated empirically in the numerical section.

Due to the sub-extensive corrections  the border of stability  $
x=0$ is expected to be only approximative for finite systems. This
implies that unstable solutions with $x>0$ and stable solutions
with $x<0$ may exist for systems of finite size. Such stable
solutions with $x<0$ and $ \Gamma>0$  are indeed found in the
following.

Finally it is pointed out that the results of subsection B are not
restricted to the thermodynamic limit and hold for systems of all
sizes. Thus the fixpoints obtained by integration of the
eqs.(\ref{9}) always correspond to stable solutions of the
modified TAP equations.
\subsection{Numerical procedure}
The numerical investigations are performed for systems up to a
size of $N=225$ with binary distributions of the bonds $
J_{ij}=\pm N^{-1/2} $ using the standard routine `NIntegrate' of
Mathematica on workstations.

The integration of the equations of motion (\ref{9}) in the regime
$x(t)>0$ is straightforward. In the region $x(t)<0$, however, a
special procedure is employed and  $(\beta\Gamma)^2$ is treated as
additional dynamical variable. The time derivative of
eq.(\ref{8neu}) yields the necessary additional equation of motion
and  the  initial value of $(\beta\Gamma)^2$ is determined from
eqs.(\ref{8neu}) for $t=0$ which finally guaranties that
eq.(\ref{8neu}) is satisfied for all times.

During the dynamic evolution $ x(t)$ generally changes  sign,
which implies a change of the numerical treatment. Therefore an
alternative, approximative  approach is partly  applied. Instead
of eq.(\ref{8neu}) the equation
\begin{equation}\label{18}
1=\frac{c}{N\,\Gamma^2}+\frac{1}{N}\sum_i\frac{\beta^2(1-m_i^2)^2}
{1+\Gamma^2\,\beta^{2}(1-m_i^2)^2}
\end{equation}
is used, where $ c $ is a small positive constant with typical
values from $ 10^{-3} $ to $ 10^{-5} $. With the supplementary
term $c/( N \Gamma^2) $ eq.(\ref{18}) has a solution $\Gamma>0$
for all values of of $x$. Due to the smallness of $ c $ only small
differences occur in comparison with the exact values of $\Gamma$,
which are $ \Gamma=0 $ for $x>0$ or which, for the case $x<0$, are
determined by eq.(\ref{8neu}). Hence the set of
eqs.(\ref{9}),(\ref{7}) and (\ref{18}) can be used for all values
of $ x(t)$ leading to an unified treatment of the cases $x<0$ and
$x>0$. No significant differences are found for the results of
these two methods.
\begin{figure}
\includegraphics[height=5cm]{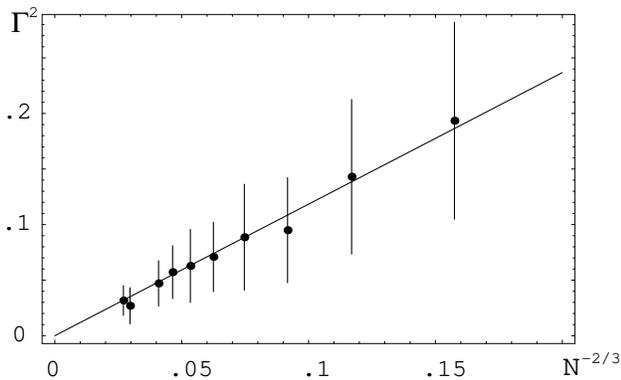}
\caption{\label{fig:f3} \textbf{ Approach to the spin glass
instability from below:} $\Gamma ^2$ versus $ N^{-2/3}$ at the
temperature T=.5. The data points represent the averages over
about $ 85$, over $ 28$ and over $25$ samples for $ N\leq 121$,
for $N=196$ and for $N=225$, respectively. The length of each
error bar is two standard deviations.}
\end{figure}

In the numerical investigations of the next section the results
are presented as functions of the temperature from $T=.01$ to
$T=1$. The temperature variation is performed both  steps-wise and
continuously. Usually the step size is $ \Delta T=.01 $. For the
case of  continuous variations the temperature is linearly  and
slowly changed. Typically  $ 10^5$ relaxation times are taken, to
change the temperature from $T= 0.1$ to $T=1$.

For runs  at a fixed temperature with $ N=100$ and with $N=225$,
several minutes and a couple of hours are needed, respectively. It
takes, however, several hours for $ N=100$  and several days for $
N=225$ when the temperature is continuously changed  in the entire
interval. At low temperature $ T<.1$ the numerical integration is
more time consuming than at higher temperatures.
\section{Numerical results}
\subsection{Finite size effects}
Let us first discuss two trivial finite size effects. The largest
eigenvalue $J_{\textrm{max}}$ of the interaction matrix $
\mathbf{J}$ determines the spin glass temperature $
T_{sg}=J_{\textrm{max}}/2$. For finite $N$ this eigenvalue
$J_{\textrm{max}}$ differs from the $ N\rightarrow\infty$ value
\cite{ma} of $2$ . The  deviations are of the order of $5\%$ and
mainly arise near $T=1$.

The second simple effect results at  low temperatures. The binary
distributions  $ J_{ij}=\pm N^{-1/2}$  imply  a splitting of the
energy levels  of $ \Delta E= 2 N^{-1/2} $. Thus for temperatures
$ T\ll 2 N^{-1/2}$ only the ground state properties enter into the
calculation of thermodynamic quantities, which causes  atypical
effects for temperature below  $T=.1 $.  These effects are not
discussed in the following \cite{cnull}.
\begin{figure}
\includegraphics[height=5cm]{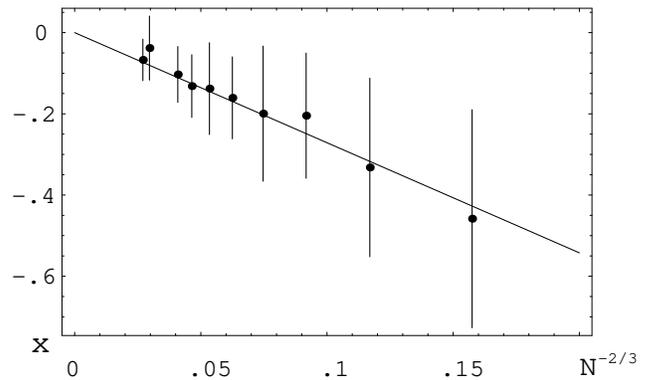}
\caption{\label{fig:f4} \textbf{ Approach to the spin glass
instability from below:} $x=1-\beta^2(1-2 q_2+q_4)$ versus $
N^{-2/3}$ at the temperature T=.5. The data points represent the
averages over about $ 85$, over $ 28$ and over $25$ samples for $
N\leq 121$, for $N=196$ and for $N=225$, respectively. The length
of each error bar is two standard deviations.}
\end{figure}

For an analysis of the convergence of the present approach with
increasing $N$  we  investigate systems of the size
$N=16,25,36,49,64,81,100, 121,196$ and  $225 $. For each size with
$N\leq121 $ one hundred different realizations  of the bonds are
investigated. For the systems with $N=196$  and with $N=225$
thirty bond configurations are examined in each case. The
temperature is fixed at $ T=.5$ and all calculations are performed
in zero field. Starting with random initial values the system
relaxes always to a fixpoint of the equation of motion that is
different from the paramagnetic solution $m_i=0$.

The fixpoint solutions are classified in two classes. In the
majority class (containing about $85 \%$ of all solutions), the
value of $ \Gamma$ is finite and the solutions are located in the
region $ x<0 $. The solutions of the minority class satisfy $x>0$
and $\Gamma=0$. Both classes are already known in literature
\cite{bm,nt} and related to the cases where solutions of the
original TAP equations are not found or are found \cite{per}.

The minority class corresponds to the  case were the TAP free
energy has a local  minimum. As shown in \cite{bm,nt} the obtained
$x $ values of these solutions tend to zero  for $
N\rightarrow\infty$ and the minimum tends to a marginally stable
saddle point. Our data are in complete agreement with these
earlier findings and need therefore no further discussion.

We focus the discussion on the majority class. All solutions of
the this class have a finite values for $\Gamma$ and satisfy $
x<0$. Both results are caused by the finite size of the systems,
as for $N\rightarrow\infty $ all solutions  of the modified TAP
equations with these properties are unstable. Thus it is expected
that both $\Gamma$ and $ x$ tend to zero when approaching  the
thermodynamic limit. Fig\ref{fig:f3} and Fig\ref{fig:f4}, where
the averaged $\Gamma^2$ and the averaged $ x$  are plotted against
$N^{-2/3}$, show indeed this expected behavior. The data are
consistent - but not conclusively - with a $ N^{-2/3}$ dependence
for the asymptotic behavior of both quantities.

From the viewpoint of the modified TAP equations both classes are
similar. The solutions of the minority class and of the majority
class describe the approach to  $x=0$ from above and from below,
respectively. As $x=0$ represents the boundary of stability in the
thermodynamic limit, this  viewpoint clearly shows the asymptotic
marginal stability of the  TAP solutions in the glassy state.
\begin{figure}
\includegraphics[height=5.3cm]{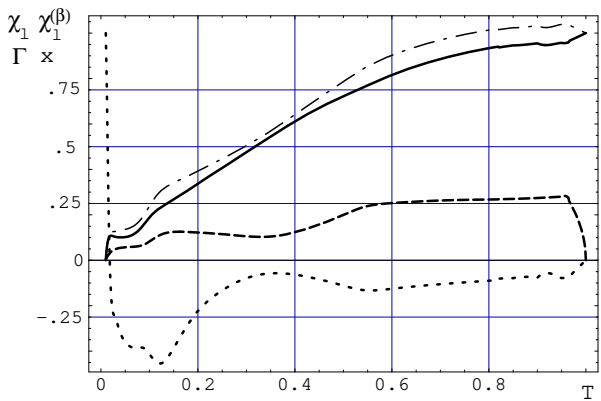}
\caption{\label{fig:f5} \textbf{Properties of the state of lowest
free energy :} Temperature dependence of $\chi_l$ (full line), of
$\chi_l^{(\beta)}=\beta(1-q_2)$ (dashed dotted line), of $\Gamma$
(dashed line) and of $x=1-\beta^2(1-2 q_2+q_4)$ (dotted line) for
a zero field cooled system with $ N=225$ (sample I) which is
representative for most systems.}
\end{figure}

Turning to temperature dependent effects we will present results
for three specific systems. From the majority class  we select
sample I and sample II with sizes  $N=225$ and  $N=100$,
respectively. Sample III  with a size of $N=100$ is chosen from
the minority class.

\begin{figure}
\includegraphics[height=5.7cm]{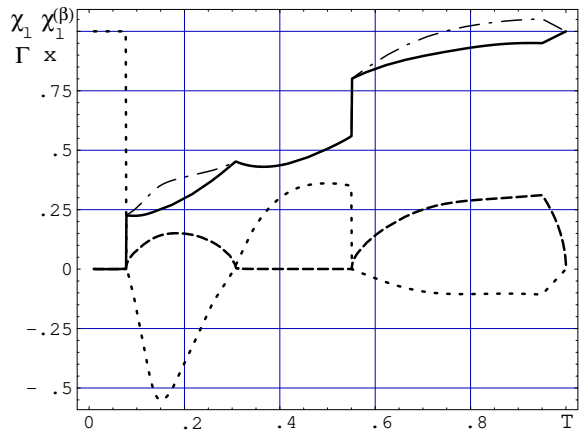}
\caption{\label{fig:f6}\textbf{ Properties of the state of lowest
free energy:} Temperature dependence of $\chi_l$ (full line), of
$\chi_l^{(\beta)}=\beta(1-q_2)$ (dashed dotted line), of $\Gamma$
(dashed line) and of $x=1-\beta^2(1-2 q_2+q_4)$ (dotted line) for
a zero field cooled system with $ N=100$ (sample III) which is
representative for only a few systems.}
\end{figure}

With the aim of finding the states with lowest free energy we
slowly cool down the systems to $ T=.01$, starting at $ T=1$ with
the paramagnetic solution as initial values. For sample I in zero
field  the numerical results for $x,\Gamma,\chi_l$ and $
\chi_l^{(\beta)}=\beta(1-q_2)$ are presented in Fig\ref{fig:f5}.
The spin glass instability is approached from below for nearly all
temperatures, where $ x<0, \:\Gamma>0 $ and $\chi_l
\neq\chi_l^{(\beta)}$ hold.  The results for sample III are
different and are plotted in Fig\ref{fig:f6}. In this sample
whether  the instability is approached from above or from below is
dependent on the temperature.

In  Fig.\ref{fig:f5} and Fig.\ref{fig:f6} discontinuities of
$\chi_l$, $\Gamma$ and $x$ arise which result from saddle node
bifurcations and which are discussed in some detail in Sec. III.D.
These jumps are expected to disappear in the thermodynamic limit.
In this case we have $x\rightarrow0$ and $\Gamma\rightarrow0 $ and
thus $\chi_l (\rightarrow\chi_l^{(\beta)})$ will become a smooth
curve. Quite remarkably the figures show that $\chi_l\rightarrow
0$ holds for $T\rightarrow0$, which is a desired behavior found in
other approaches \cite{mpv,fh}.

The procedure of slowly cooling down to get the states of lowest
free energy can at best be justified by empirical arguments.
Indeed,  further states of the samples under investigation were
found to  have in general, but not always  a higher free energy
(compare Sec. III.C.).

To obtain some more evidence for this procedure, we  apply  our
approach to systems with zero fields and with $ N=10,11,\ldots 22$
and  investigate for each $N$ one hundred sets of the bonds. In
each case the state of lowest free energy at $ T=0$ is determined
by slowly cooling down from $T=1$ and compared with the exact
result which can be calculated due to the smallness of the
systems. We obtain agreement in $95\%$ of all cases and in a
further $ 4\%$ of the cases the cooling method found one of the,
in general degenerate, lowest exited states. A significant
dependence on $N$ is not found. Provided that an extrapolation to
large $ N$ is possible, these results are quite remarkable, even
from the viewpoint of optimization problems.
\subsection{ Properties of the state of lowest free energy}
\begin{figure}
\includegraphics[height=5cm]{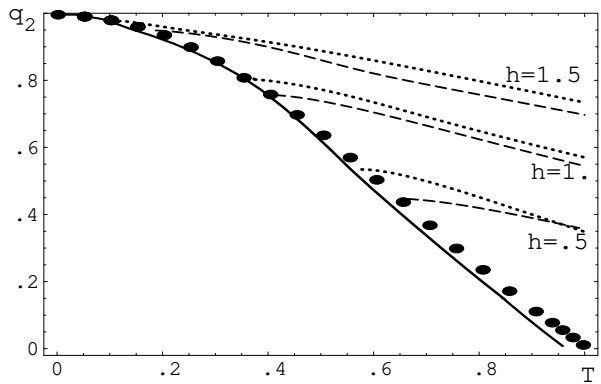}
\caption{\label{fig:f7}\textbf{Edwards Anderson order parameter of
the state of lowest free energy :}  $q_2$ versus temperature $T$
in zero field (full line) and in external fields $h$ for sample I
with $N=225$ (dotted lines) and for sample II with $N=100$ (dashed
lines). At the temperatures where the lines for finite fields end
the glassy regime is entered  and $x $ becomes negative. The dots
represent the results of the replica approach \cite{cr} for $ h=0
$.}
\end{figure}
\begin{figure}
\includegraphics[height=5cm]{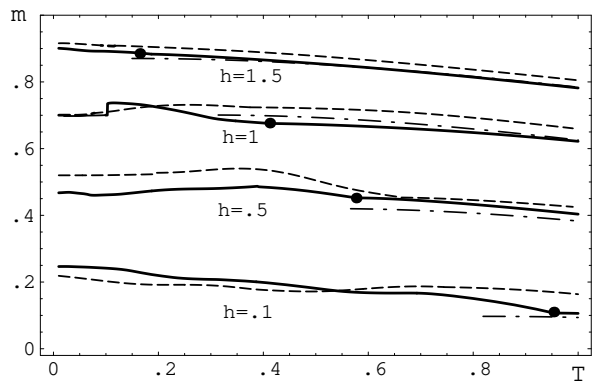}
\caption{\label{fig:f8} \textbf{ Total magnetization of the state
of lowest free energy :}  $ m=N^{-1}\sum m_i $ versus temperature
in external fields $h$ for sample I with $N=225$ (full lines) and
for sample II with $N=100$ (dashed lines). The dashed dotted line
shows the SK results ending at the AT temperature and the dots
mark the temperatures where $ x $ changes sign.}
\end{figure}
\begin{figure}
\includegraphics[height=6cm]{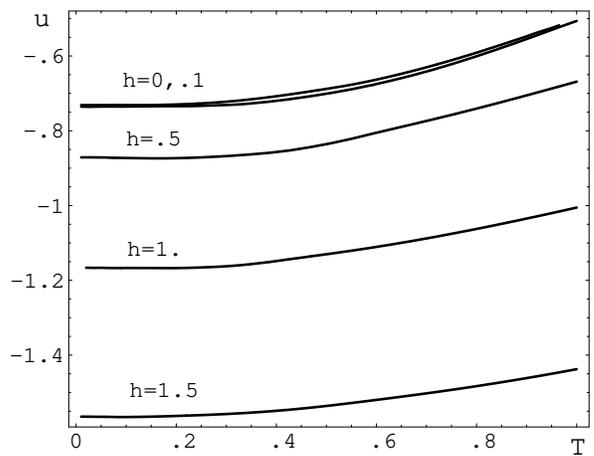}
\caption{\label{fig:f9} \textbf{Energy density of the state of
lowest free energy:}  $u=U/N$ versus temperature $T$ in external
fields $h$ for sample I with $N=225$.}
\end{figure}
\begin{figure}
\includegraphics[height=5.8cm]{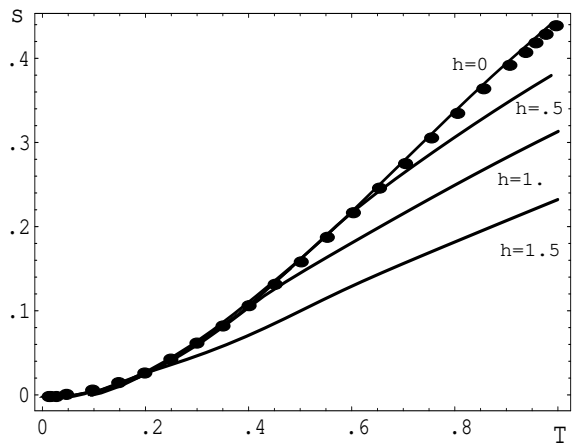}
\caption{\label{fig:f10} \textbf{Entropy density of the state of
lowest free energy:}  $s=S/N$ versus temperature $T$ in external
fields $h$ for sample I with $N=225$.  The dots represent the
results of the replica approach \cite{cr} for $ h=0$ .}
\end{figure}
In this subsection the results of various physical quantities will
be presented for all temperatures below the spin glass temperature
in the presence of a homogeneous external field $h$. In some cases
the results will exclusively be given for sample I and in other
cases the results for sample II will be added to visualize the
finite size effects.

In Fig.\ref{fig:f7} the Edwards Anderson order parameter $q_2$ is
plotted versus $T$. The finite size effects are rather large in
the paramagnetic regime, but are reasonable in the spin glass
regime where  $q_2 $ is nearly independent of the field $ h$. The
overall behavior of $q_2$ is in agreement with \cite{ptv}, with
the numerical studies of \cite{nt} and  with the replica approach
\cite{cr}. Note that the agreement with the results \cite{cr}
could be improved by  rescaling the temperature axis to identical
spin glass temperatures.

In Fig.\ref{fig:f8} the  $T$ and $h$ dependence of the total
homogenous magnetization $m=N^{-1}\sum_i m_i$  is presented. This
figure  also shows the result of the replica theory \cite{sk}
above the AT line \cite{at} being exact in the thermodynamic
limit. Again the finite size effects are relatively large. The
general behavior, however, can  clearly be identified.

The energy density $ u=U/N $ and the entropy density $ s=S/N$ are
plotted in Fig.\ref{fig:f9} and in Fig.\ref{fig:f10},
respectively. Only the results for sample I are given, as the
differences to sample II are not resolved  on the scale of these
figures. For the calculation of the plotted quantities
eq.(\ref{81}) and eq.(\ref{82}) are used. This use implies again
an approximation, as strictly these asymptotic equations hold only
for $\Gamma=0$. Note finally that the results of Fig.\ref{fig:f10}
are again in agreement to the results  of \cite{cr} based on the
replica approach.
\subsection{Meta stable  states}
\begin{figure}
\includegraphics[height=4.8cm]{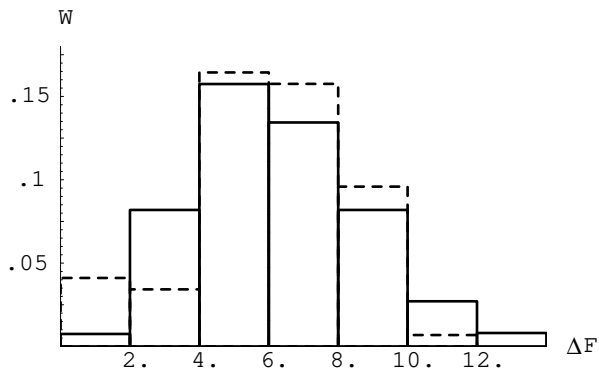}
\caption{\label{fig:f14}\textbf{Meta stable states:} Distribution
$W $ of the excess free energy $\Delta F=F-F^0$ at $T=.2$ in zero
field for sample I (full line) and for sample II (dashed line).}
\end{figure}
Let us  estimate  the expected number of solutions $ N_s$ of the
TAP equations for our system sizes. At $T=.2$,  according to Bray
and Moore \cite{bm83}, this number is approximately given by $
N_s\approx \exp(.05 \,N) $. In zero magnetic field for each
solution a further solution can trivially be constructed by
changing the sign of all $ m_i$. Counting each pair as one
independent solution the  estimates for the number of independent
solutions are $ 75 $ and $ 3850 $  for $N=100$ and $N=225$,
respectively. Note that the basis for this estimate are the
original and not the modified TAP equations and thus these numbers
should be used with some care for the present work.

The investigations for finding  meta stable sates are performed in
zero external field at $ T=.2$, starting each run with different
random initial values of the $ m_i$.

In $ 540 $ runs  we have found $74 $ independent solutions of the
modified TAP equations for sample II. In the first $100$ runs
nearly one half of them are obtained. Afterwards the rate for
finding new solutions quickly decreases to  about $ 8 \%$ and
remains approximately constant for later runs. Thus certainly not
 all existing solutions of this sample are observed. The
statistical results, presented in the following, are nearly
independent of the solutions of later runs. This implies that the
observed solutions are representative for all solutions of this
sample. A further argument for the last statement results from the
fact that the states with the $8$ lowest free energies are found
in the first $100$ runs. This preference for the lower states is
also found in the rate of recurrence of all runs, which indicates
a large basin of attraction for these states.

For sample I with $ N=225 $ we have calculated $ 868 $ different
states in $ 1250 $ runs. For this sample the rate for finding new
solutions decreases  only to a value of about $ 50 \%$.
Nevertheless it is assumed that the  solutions found are again
representative for all solutions. In this context it should be
added that about $ 8\%$ of the meta stable states of both samples
have a $x> 0$ and belong therefore to the minority class.

Fig.\ref{fig:f14} shows the normalized distribution $ W( \Delta F)
$ of the excess free energies  $ \Delta F^\alpha= F^\alpha-F^0$ of
the meta stable states, where $ F^\alpha$ and $ F^0$ denotes the
free energies of the meta stable states and of the state of lowest
free energy, respectively. Note that the range of $ \Delta
F^\alpha $ is identical for both samples and seems to be
independent of $N$. Moreover even the distribution itself seems to
be independent of $N$.

Assuming that the findings can be  generalized   this is an
interesting result for the `multi-valley structure' of the free
energy. It implies that all the  possible excess free energies $
\Delta F^\alpha $ remain finite in the thermodynamic limit. Note
that this conclusion is in agreement with the result of \cite{II}
that the `multi-valley' structure of the TAP free energy occurs on
a sub-extensive scale.
\begin{figure}
\includegraphics[height=5.0cm]{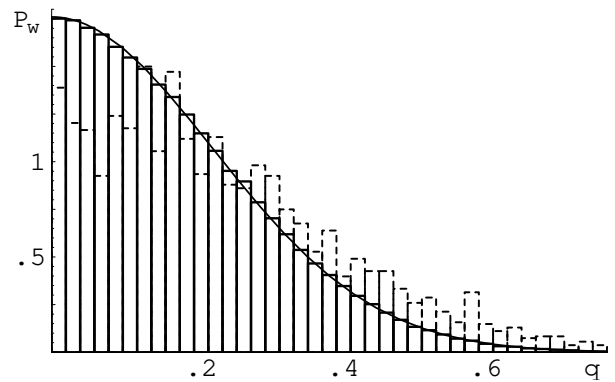}
 \caption{\label{fig:f13} \textbf{Meta stable states:}
White-weighted overlap distribution at $T=.2$ in zero field for
sample I (full line) and for sample II (dashed line). The smooth
line shows a Gaussian fit of the data of sample I.}
\end{figure}
\begin{figure}
\includegraphics[height=10cm]{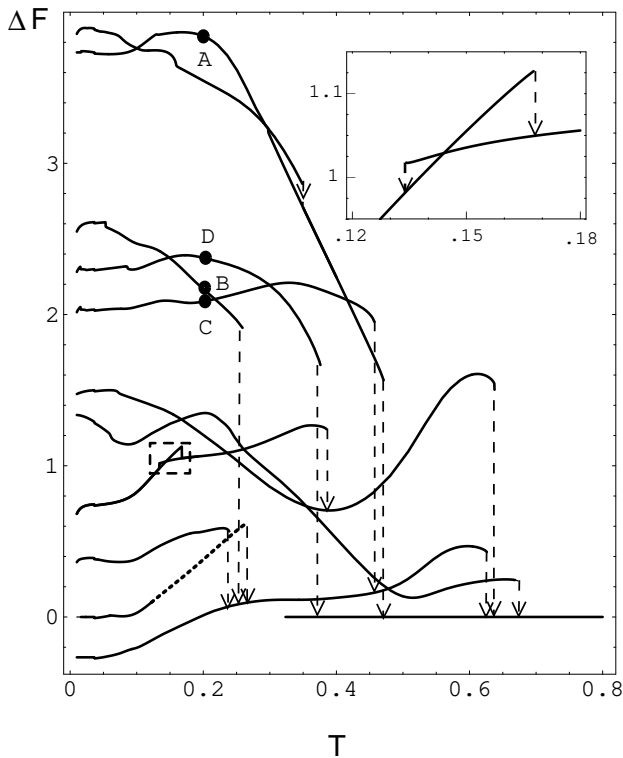}
\caption{\label{fig:f12}\textbf{Meta stable states:} Excess free
energy  $ \Delta F $ of sample I versus temperature $T$ of the six
lowest and four further meta stable states which can be reached by
physical processes (compare text).The full and the dotted lines
correspond to $x<0$ and $x>0$, respectively and the dashed arrows
indicate the observed jumps when changing  $T$ . The insert shows
the details of a hysteresis near a discontinuity.}
\end{figure}

Let us introduce the overlap $ q^{\alpha \alpha'} $ between the
solution $m_i^\alpha$ and the solution $m_i^{\alpha'}$  in the
usual way
\begin{equation}\label{21}
  q^{\alpha \alpha'}=N^{-1}\sum_i m_i^\alpha m_i^{\alpha'}
\end{equation}
and define the normalized, white-weighted overlap distribution
\begin{equation}\label{22}
  P_w(q)=\frac{2}{\hat{N}_s(\hat{N}_s-1)} \sum_{\alpha<\alpha'} \delta(q-q^{\alpha
  \alpha'}).
\end{equation}
where $\hat{N}_s$ denotes the total number of solutions. (Note
that $P_w(q)$ is not the Parisi overlap distribution, as the
Boltzmann weights are absent.)

The distribution $P_w(q)$ is symmetrical  in $q$ and is plotted
for $q>0$ as histogram in Fig.\ref{fig:f13} for samples I and for
sample II. Both distributions are similar and thus nearly
independent of the system size $N$.  The resulting asymptotic
distribution seems to be a Gaussian distribution. Such a
distribution  would result if all the directions of the meta
stable states in the $ N$ dimensional space  are not correlated.

Unfortunately the data for the meta stable states are not
sufficient to answer the interesting question of whether the
Parisi overlap distribution, a quantity averaged over all bonds $
J_{ij}$, applies also to single systems. Both samples have just
three states with non negligible Boltzmann weights and thus
definitive conclusions are impossible.

The temperature dependence of the meta stable states is presented
for (at $T=.2$) the lowest six and some higher states of sample I
in Fig.\ref{fig:f12}. The branches of the latter states  can be
reached by physical processes. These processes consist of a slow
field cooling part which starts at $T=1$ and which is followed by
a sudden switching-off  of the field at $T=.2$. For the  values
$h=1.5$, $h=1.0$, $h=.5$  and $h=.1$ the processes end at the
points A,B,C and D of Fig.\ref{fig:f12}, respectively.

All the meta stable states investigated vary smoothly in some
temperature region which is limited by an upper limiting
temperature. Above this temperature  the states disappear via
saddle node bifurcations. For some states an additional  lower
limiting temperature is found (compare insert of
Fig.\ref{fig:f12}). The system remains in a definite meta stable
state for all slow changes of the temperature as long as the
limiting temperatures are not reached. However approaching  a
limiting temperature by slow temperature changes, the system
always bifurcates into a branch which has a lower free energy,
implying  hysteresis effects in some cases. Although such results
are physically expected this behavior is numerically verified for
all states plotted in Fig.\ref{fig:f12}.

Note that in the whole temperature range most of the meta stable
states have negligible Boltzmann weights. Actually only the three
lowest states of sample I will give non negligible contributions
to Boltzmann averaged quantities. Nevertheless the system will
remain  dynamically in a meta stable state of high free energy for
all times provided it remains there initially and provided no
drastic changes of the temperature or the magnetic field are
performed.

According to Fig.\ref{fig:f12} the free energy levels sometimes
cross with changes of the temperature. Even the state of lowest
free energy can be involved, which implies that  the state of
lowest free energy and the first exited state change their roles
at the crossing temperature. Thus in the notation of this work, to
call the state slowly cooled from $T=1$ the state of lowest free
energy is  not strictly correct. For brevity  we will still use
this term keeping, however, the limitations in mind.
\subsection{Saddle node bifurcations}
In this subsection the discontinuities are analyzed which arise in
the temperature dependence of states of lowest free energy and of
the meta stable states.

\begin{figure}
\includegraphics[height=7cm]{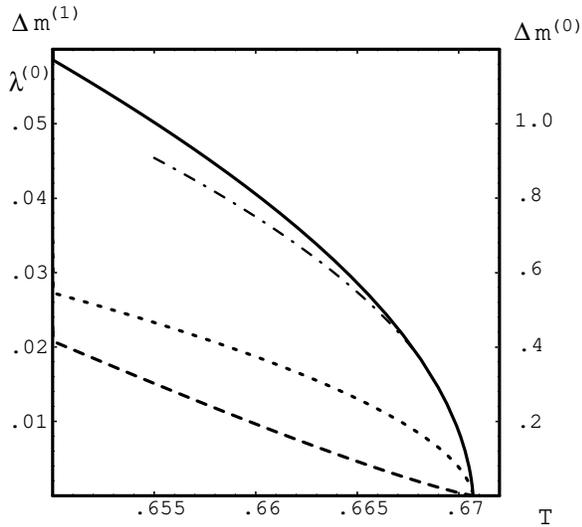}
\caption{\label{fig:s} \textbf{ Saddle node bifurcation:} Smallest
eigenvalue $ \lambda^{(0)} $ (dotted line), critical staggered
magnetization  $\Delta m^{(0)}$ (full line) and an uncritical
staggered magnetization  $\Delta m^{(1)}$ (dashed line) versus
temperature $T$ near the discontinuity at $ T_c= .6707$ of
Fig.\ref{fig:f12}. The dashed dotted line shows the leading
behavior $\Delta m^{(0)}\sim ( T_c-T)^{.5} $ near $T_c$.}
\end{figure}

For this purpose the eigenvalues $
\lambda^{(0)}\leq\lambda^{(1)}\leq\lambda^{(2)}\ldots$ and the
eigenvectors $ u^{(\alpha)} $ of  the inverse susceptibility
(\ref{13})
\begin{equation}\label{40}
\bchi^{-1}u^{(\alpha)}=\lambda^{(\alpha)}u^{(\alpha)}\quad
\alpha=0,1\ldots N-1.
\end{equation}
are introduced for an arbitrary  solution $m_i$ at the temperature
$T$. Approaching a discontinuity the smallest eigenvalue $
\lambda^{(0)}$ tends to zero and vanishes at a critical
temperature $ T_c$. In Fig.\ref{fig:s} an example is presented
where the temperature dependence of  $ \lambda^{(0)}$ is given by
the numerical  solutions of eq.(\ref{40}).

For  further discussions the staggered magnetizations $ \Delta
m^{(\alpha)} $ are defined
\begin{equation}\label{41}
\Delta m^{(\alpha)}=\sum_i{\big\{}m_i(T)-m_i(T_c){\big\}}\;
u_i^{(\alpha)}(T_c)\quad \alpha=0,1\ldots
\end{equation}
The numerical results for $\Delta m^{(0)}$ and $\Delta m^{(1)}$
are plotted  in Fig.\ref{fig:s} for a discontinuity. Within the
numerical errors $\Delta m^{(0)}\sim (T_c-T)^{.5}$ and $ \Delta
m^{(1)}\sim (T_c-T )$ hold where the latter result is
representative for all $\alpha\neq0$.

These temperature variations  can analytically be explained. For
such an approach eq.(\ref{4}) is rewritten as
\begin{equation}\label{42}
h_i(T,m_k)= \frac{T}{2} \ln \frac{1+m_i}{1-m_i}-\sum_j
J_{ij}m_j+m_i \chi_l
\end{equation}
and expanded  at $ T_c$ to first order in $ \Delta T=T_c-T$ and to
second order in $\Delta m_i= m_i(T)-m_i(T_c)$ which yields
\begin{equation}\label{43}
A_i\Delta T= \sum_j\chi^{-1}_{ij}\Delta m_j +
\sum_{jk}\frac{B_{ijk}}{2}\Delta m_j\Delta m_k
\end{equation}
with
\begin{equation}\label{44}
A_i= \partial_ T \;h_i (T,m_k)\big|_{T_c}\;\mathrm{and}\;B_{ijk}=
\partial_{m_k} \;\chi_{ij}^{-1} (T,m_k)\big|_{T_c}\;.
\end{equation}
Multiplying eq.(\ref{43}) from the left by  the eigenvectors
$u^{(\alpha)}(T_c)$ the $\Delta m^{(\alpha)}$  are calculated to
\begin{equation}\label{45}
\Delta m^{(0)}=\pm \sqrt{2 A^{(0)}/B^{(0)}\Delta T}\;+\;O(\Delta T
^{3/2})
\end{equation}
and to
\begin{equation}\label{46}
\Delta m^{(\alpha\neq 0)}=
\frac{A^{(\alpha)}B^{(0)}-A^{(0)}B^{(\alpha)}}{B^{(0)}\lambda^{(\alpha)}(T_c)}
\Delta T \;+\;O(\Delta T^{3/2})
\end{equation}
where
\begin{equation}\label{47}
A^{(\alpha)}=\sum_i  u_i^{(\alpha)}(T_c)\;A_i
\end{equation}
and
\begin{equation}\label{48}
B^{(\alpha)}=\sum_{ijk}
u_i^{(\alpha)}(T_c)\;B_{ijk}\;u_j^{(0)}(T_c)\,u_k^{(0)}(T_c)\;.
\end{equation}
The analytic results (\ref{45}) and (\ref{45}) agree with
temperature dependence found numerically and plotted in
Fig.\ref{fig:s}. Moreover these results exhibit all the typical
features of a saddle-node bifurcation. The two branches exist only
for $\Delta T >0$ or for $\Delta T <0$ depending on the  signs of
$A^{(0)}$ and of $B^{(0)}$. Provided the branches exist one of
them is stable and the other is unstable \cite{comx}.

The staggered magnetizations $ \Delta m^{(0)}$ govern the
behaviors of the states near the critical temperatures. Thus the $
\Delta m^{(0)}$ \textit{are the order parameters of the local
saddle node bifurcations}.

\begin{figure}
\includegraphics[height=6.5cm]{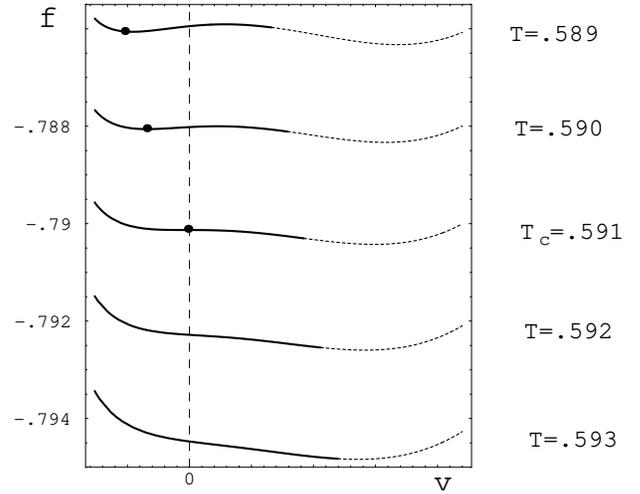}
\caption{\label{fig:a4} \textbf{Free energy landscape:} Free
energy density $ f=F/N $  versus distance $v$ from the critical
magnetization in direction of $u^{(0)}$  for temperatures above
and below the critical temperature $ T_c=.591$ (compare text).
Shown is the heating process of sample III for the state of lowest
free energy. The stable solutions are indicated by dots. The full
and the dashed lines correspond to the regimes $x>0$ and $x<0$,
respectively.}
\end{figure}
Further numerical  evidence for such bifurcations is presented in
Fig.\ref{fig:a4} where the discontinuity of sample III at $
T=.591$ for the heating process is considered. In this figure the
temperature variation of the free energy landscape is investigated
by plotting the quantity
\begin{equation}
f(T,v)= N^{-1} F(T,m_i\rightarrow m_i(T_c) +v \;u^{(0)}_i(T_c)).
\end{equation}
Below a critical temperature a minimum of the free energy and
hence a stable solution exists in the region $ x>0$ which
disappears by a saddle-node bifurcation above the critical
temperature. Note that the stationary values of $v$ determine
approximately the temperature dependence of $F$ near $T_c$.

A further interesting detail is contained in Fig.\ref{fig:a4}.
Below the critical temperature $ f$ is not semi-convex everywhere
for $ x>0$, but above the critical temperature $ f $ is convex
even in some regime $ x<0$. This demonstrates explicitly that the
border of stability $x=0$ is only approximative for finite
systems.
\section{Conclusions}
In this work we have demonstrated that the modified TAP equations
are an adequate tool for exploring the characteristics of the pure
states of the SK model  of finite sizes. For all realizations of
the random bonds explicit solutions exhibiting the spin glass
features are obtained. Moreover, the present approach makes it
possible to analyze  the  temperature dependence of all quantities
of physical interest.  In contrast to other approaches the method
presented can simulate a slow cooling  of the system, which is of
particular  importance to find the features of the low lying
states.

Apart from these properties for the state of lowest free energy we
have found some interesting statistical features of the meta
stable states which seems to be generic for every sample and in
the author's opinion, it is a challenge to find an analytical
approach to these features.

Due to finite size effects most of the results of this work are
only qualitative.  Thus  systems  of larger size should be
investigated. This seems to be at least partially possible.

The present work uses a relaxational dynamics as a mathematical
tool to find the pure states  as fixpoints of the equations of
motion. Recall that the origin  of this dynamics is
phenomenological and thus true dynamic effects cannot  be
described by the present approach. Thus the complete dynamical
description should be a subject of further research.


\begin{thebibliography}{20}
\bibitem{cnull}  These arguments do
not apply exactly at $T=0$. Thus quantities, like the  ground
state energy, its degeneration and the ground state spin
configuration are  typical properties.
\bibitem{per} Note that solutions found in this investigation may be
meta stable states. This is indeed the case for most members of
the minority class and the rate of finding a solution of the
minority class which like sample III represents  a state of lowest
free energy is significantly lower than $15 \%$.
\bibitem{comx}
Standard first order perturbation theory leads to $
\lambda^{(0)}=B^{(0)} \Delta m^{(0)} + O(\Delta T)$ which implies
the claimed result. Moreover  $ \lambda^{(0)}\sim\sqrt{ |\Delta
T|} $  holds for the stable branch in aggreement with
Fig.\ref{fig:s}.
\end{thebibliography}
\end{document}